\documentclass[showpacs,preprintnumbers,amsmath,amssymb,twocolumn]{revtex4}%

\usepackage{graphicx}
\usepackage{dcolumn}
\usepackage{bm}
\usepackage{color}

\begin{document}


\title{Particle Motion around Black Hole in
Ho\v{r}ava-Lifshitz Gravity}

\author{Ahmadjon Abdujabbarov}%
   \email{ahmadjon@astrin.uz}

\author{Bobomurat Ahmedov}
 \email{ahmedov@astrin.uz}

\author{Abdullo Hakimov}
   \email{abdullo@astrin.uz}

 \affiliation{%
Institute of Nuclear Physics, Ulughbek, Tashkent 100214,
Uzbekistan\\
Ulugh Beg Astronomical Institute,    Astronomicheskaya 33,
Tashkent  100052, Uzbekistan
    }

\date{\today}
\begin{abstract}

Analytical solutions of Maxwell equations around black hole
immersed in external uniform magnetic field in the background of
the Kehagias-Sfetsos (KS) asymptotically flat black hole solution
of Ho\v{r}ava-Lifshitz gravity have been found. Influence of
magnetic field on effective potential of the radial motion of
charged test particle around black hole immersed in external
magnetic field in Ho\v{r}ava-Lifshitz gravity has been
investigated by using Hamilton-Jacobi method. Exact analytical
solution for dependence of the minimal radius  of the circular
orbits $r_{\rm mc}$ from KS parameter $\omega$ for motion of test
particle around spherical symmetric black hole in
Ho\v{r}ava-Lifshitz gravity has been derived. The critical values
of the particle's angular momentum for captured particles by black
hole in Ho\v{r}ava-Lifshitz gravity have been obtained
numerically. The comparison of the obtained numerical results with
the astrophysical observational data on radii of the innermost
stable circular orbits gives us the estimation of the parameter as
$\omega \simeq 3.6 \cdot 10^{-24} {\rm cm}^{-2}$.

\end{abstract}

\maketitle

\section{\label{sec:intro}Introduction}

Recently Petr Ho\v{r}ava suggested new candidate quantum field
theory of gravity with dynamical critical exponent equal to $z =
3$ in the UV (Ultra-Violet). This theory is a non-relativistic
power-counting renormalizable theory in four dimensions, which
admits the Lifshitz scale-invariance in time and space that
reduces to Einstein's general relativity at large scales
\cite{h1}, \cite{h2}. The Ho\v{r}ava theory has received a great
deal of attention and since its formulation various properties and
characteristics have been extensively analyzed, ranging from
formal developments \cite{Visser}, cosmology \cite{Takashi}, dark
energy \cite{Saridakis}, dark matter \cite{Mukohyama}, and
spherically symmetric or axial symmetric solutions \cite{CaiCao}.

In the paper \cite{lobo1} the possibility of observationally
testing Ho\v{r}ava gravity at the scale of the Solar System, by
considering the classical tests of general relativity (perihelion
precession of the planet Mercury, deflection of light by the Sun
and the radar echo delay) for the Kehagias-Sfetsos (KS)
asymptotically flat black hole solution of Horava-Lifshitz gravity
has been considered. The stability of the Einstein static universe
by considering linear homogeneous perturbations in the context of
an Infra-Red (IR) modification of Ho\v{r}ava gravity has been
studied in~\cite{lobo2}. Potentially observable properties of
black holes in the deformed Ho\v{r}ava-Lifshitz gravity with
Minkowski vacuum: the gravitational lensing and quasinormal modes
have been syudied in~\cite{konoplya11}. {The authors of the paper
\cite{einstein} derived the full set of equations of motion, and
then obtained spherically symmetric solutions for UV completed
theory of Einstein proposed by Ho\v{r}ava. }

{The paper is organized as follows.} We look for exact solutions
of vacuum Maxwell equations in spacetime of the black hole
immersed in uniform magnetic field in IR modified
Ho\v{r}ava-Lifshitz gravity in Sect.~\ref{sec:2}. In our recent
paper~\cite{aaprd} exact analytical solution for dependence of the
radius of the innermost stable circular orbits (ISCO) $r_{\rm
ISCO}$ from brane tension for motion of test particle around black
hole in the braneworld has been analyzed. We extend it to the
motion of charged particles around black hole immersed in uniform
magnetic field in Ho\v{r}ava-Lifshitz gravity using the
Hamilton-Jacobi method in Sect.~\ref{sec:3}. We obtain the
effective potential for charged test particle with a specific
angular momentum, orbiting around the black hole, as a function of
the external magnetic field, and the IR-modified parameter in
Ho\v{r}ava-Lifshitz gravity. In Sect.~\ref{sec:4} we find the
exact expression for dependence of minimal radius of circular
orbit from parameter $\omega$, which is responsible for the IR
modified term in the Ho\v{r}ava-Lifshitz action, for the test
particle moving in the equatorial plane of the black hole(when the
external magnetic field is neglected for the simplicity of
calculations). Then we present the numerical results of the
capture cross section of the slowly moving test particles by the
black hole in Ho\v{r}ava-Lifshitz gravity. {The concluding remarks
are given in Sect.~\ref{conclusion}.}

We use a system of units in which $c = G = 1$, a space-like
signature $(-,+,+,+)$ and a spherical coordinate system
$(t,r,\theta ,\varphi)$. Greek indices are taken to run from 0 to
3. We will indicate vectors with bold symbols ({\it e.g.} $\mathbf
B$) .

\section{\label{sec:2}  Black Hole Immersed in Uniform Magnetic Field}

The static and spherical symmetric spacetime metric of the black
hole with mass $M$ in Ho\v{r}ava-Lifshitz gravity  takes form
{(see e.g, \cite{lobo1,lobo2,konoplya11})}
\begin{equation}\label{metric}
ds^2=-e^{2\Phi(r)}dt^2+e^{2\Lambda(r)}dr^2 +r^2d\theta^2+r^2
\sin^2 \theta d\varphi^2 \ ,\ \quad
\end{equation}
where the metric functions $\Phi$ and $\Lambda$ depend on the
radial coordinate $r$ only.

We consider the Kehagias and Sfetsos's  asymptotically flat
solution \cite{ks09}, given by
\begin{equation}
e^{2\Phi(r)}=e^{-2\Lambda(r)}= 1+\omega r^2-\sqrt{r(\omega^2
r^3+4\omega M)}\label{lapse}.
\end{equation}
%


{A Killing vector $\xi^\mu$ being an infinitesimal generator of an
isometry satisfies to the equation
\begin{equation}\label{2.1}
 \xi_{\alpha ;\beta}+\xi_{\beta;\alpha}=0\ ,
\end{equation}
which can be used in order to rewrite the equation
\begin{equation}
\xi_{\alpha;\beta;\gamma}-\xi_{\alpha;\gamma;\beta}=-\xi^\lambda
R_{\lambda\alpha\beta\gamma}\  ,
\end{equation}
which defines the Riemann curvature tensor
$R_{\lambda\alpha\beta\gamma}$ in the form
\begin{equation}\label{ricci}
\xi^{\alpha;\beta}_{\ \ \ ;\beta}=\xi^\gamma R_{\gamma\beta}^{\ \
\ \alpha\beta}=R^{\alpha}_{\ \gamma} \xi^{\gamma}\ .
\end{equation}
}

For spacetime of the KS black hole the right hand side of the
equation (\ref{ricci}) can be expressed as $R^{\alpha}_{\ \gamma}
\xi^{\gamma}=\eta^{\alpha}$ and consequently the Maxwell equations
as
\begin{equation}\label{waldself}
F^{\alpha\beta} _{\ \ \ ;\beta}=-2C_0\left(\xi^{\alpha;\beta} _{\
\ \ ;\beta} -  \eta^{\alpha}\right)=0\ ,
\end{equation}
where $\eta^{\alpha}=\left\{0,0,0,6 M^{2}/ \omega r^{6}\right\}$
{is just the first order approximation in $\omega^{-1}$ of the
relation $R^{\alpha}_{\ \gamma}\xi^{\gamma}$(see \cite{ks09}) and
neglecting the time component $\eta^{t}$ which can be explained as
follows. If one considers the electrical neutrality of the source:
\begin{eqnarray}
\label{flux}  && 4\pi Q=0= \frac{1}{2}\oint
F^{\alpha\beta}{_*dS}_{\alpha\beta}\  ,\nonumber
\end{eqnarray}
where ${_*dS}_{\alpha\beta}$ is the element of a 2-surface, and
evaluate the value of the integral through the spherical surface
at the asymptotic infinity ($r\rightarrow \infty, \ \ \omega r^2
\rightarrow\infty$), one can obtain that time component of the
potential will vanish identically (see, for more details
\cite{aak08}). Taking into the account the Lorentz gauge} the
electromagnetic field tensor $F_{\alpha\beta}$ can be selected as
\begin{eqnarray}
F_{\alpha\beta}&=&C_0\left(\xi_{\beta;\alpha}-\xi_{\alpha
;\beta}+2f_{\alpha\beta}\right)=\nonumber\\&&-2C_0\left(\xi_{\alpha
;\beta}+a_{\beta,\alpha}-a_{\alpha,\beta}\right)\ .
\end{eqnarray}
Here $C_0$ is constant and 4-potential $a^\alpha$ being
responsible for the KS parameter $\omega$ can be found from the
equation $ \Box{a^\alpha}=\eta^{\alpha}$.

{Finally one can express the electromagnetic potential as a sum of
two contributions}
\begin{equation}
A^\alpha=\tilde{A}^\alpha+ a^\alpha\ ,
\end{equation}
{where $\tilde{A}^\alpha$ is the potential being proportional to
the Killing vectors. To find the solution for $\tilde{A}^\alpha $
we exploit the existence in this spacetime of a timelike Killing
vector $\xi^\alpha_{(t)}$ and spacelike one
$\xi^\alpha_{(\varphi)}$ being responsible for stationarity and
axial symmetry of geometry, such that they {satisfy the} Killing
equations (\ref{2.1}) and {consequently the} wave-like equations
(in vacuum spacetime)
$ \Box{\xi^\alpha}=0\ , $
which gives a right to write the solution of vacuum Maxwell
equations $\Box \tilde{A}^\alpha=0$ for the vector potential
$\tilde{A}_\alpha$ of the electromagnetic field in the Lorentz
gauge in the simple form
$
 \tilde{A}^\alpha=C_1 \xi^\alpha_{(t)}+C_2
\xi^\alpha_{(\varphi)}\ $ \cite{wald}.
The constant $C_2=B/2$, where gravitational source is immersed in
the uniform magnetic field $\textbf{B}$ being parallel to its axis
of rotation. The value of the remaining constant $C_1$ will vanish
which can be easily shown from the asymptotic properties of
spacetime (\ref{metric}) at the infinity (see e.g. our preceding
papers~\cite{aaprd,aak08} for the details of typical
calculations).}

{The second part $a^\alpha$ of the total vector potential of
electromagnetic field is produced by the presence of the KS
parameter $\omega$ and has the following solution
$$a^\alpha=\frac{B}{2}\left\{0,0,0,\frac{3M^2}{10 {\omega} r^4}\right\}\ .$$
 }

Finally the 4-vector potential $A_\alpha$ of the electromagnetic
field will take a form
\begin{eqnarray}
\label{potential} A_0=A_1=A_2=0\ , \ \ \
A_3=\frac{1}{2}Br^2\sin^2\theta \left(1+\frac{3M^2}{10{\omega}
r^4}\right).
\end{eqnarray}
%

The orthonormal components of the electromagnetic fields measured
by fixed observer with the four-velocity components
$ (u^{\alpha})_{\rm obs}\equiv \exp\left(-\Phi\right)\{1,0,0,0\}$
; $(u_{\alpha})_{\rm obs}\equiv-\exp\left(\Phi\right)\{1,0,0,0\} $
are given by expressions
\begin{eqnarray}
&&\label{magnetr} B^{\hat r} = B \left(1+\frac{3M^2}{10{\omega}
r^4}\right) \cos\theta , \ \\
&& \label{magnett} B^{\hat\theta} = e^{\Phi(r)} B
\left(1-\frac{6M^2}{10{\omega} r^4}\right) \sin\theta ,
\end{eqnarray}
which depend on the lapse function of the metric~(\ref{metric}).

In the limit of flat spacetime, i.e. for $\omega r^2\rightarrow
\infty$ and $2M/r\rightarrow 0$, expressions (\ref{magnetr}) --
(\ref{magnett}) gives
\begin{eqnarray}
 && \label{limit_B_1} \lim_{\omega r^2\rightarrow \infty ,
2M/r\rightarrow 0} B^{\hat r}=B\cos\theta
    \ , \\ \label{limit_B_2} &&
 \lim_{\omega r^2\rightarrow \infty ,
2M/r\rightarrow 0} B^{\hat\theta}=B\sin\theta
     \ .
\end{eqnarray}
 As expected, expressions
(\ref{limit_B_1}) -- (\ref{limit_B_2}) coincides with the
solutions for the homogeneous magnetic field in the Newtonian
spacetime.


\section{\label{sec:3}Motion of the Charged Particles Around Black
Hole}


It is very important to study in detail the motion of charged
particles around a black hole in Ho\v{r}ava-Lifshitz gravity
immersed in a uniform magnetic field  given by 4-vector potential
(\ref{potential}) with the aim to find astrophysical evidence for
the existence of KS parameter $\omega$.

We shall study the motion of the charged test particles around
black hole in Ho\v{r}ava-Lifshitz gravity using the
Hamilton-Jacobi equation
\begin{equation}
\label{Ham-Jac-eq} g^{\mu\nu}\left(\frac{\partial S}{\partial
x^\mu}+eA_\mu\right)\left(\frac{\partial S}{\partial
x^\nu}+eA_\nu\right)=-m^2\ ,
\end{equation}
where $e$ and $m$ are the charge and the mass of a test particle,
respectively. Since $t$ and $\varphi$  are  the Killing variables
one can write the action in the form
\begin{equation}
S=-{\cal E}t+{\cal L}\varphi+S_{\rm r\theta}(r,\theta)\ ,
\end{equation}
where the conserved quantities $\cal E$ and $\cal L$ are the
energy and the angular momentum of a test particle at infinity.
Substituting it into equation (\ref{Ham-Jac-eq}) one can get the
equation for inseparable part of the action
\begin{eqnarray}
 -m^2 &=& -e^{2\Lambda(r)}{\cal E}^2+e^{2\Phi(r)}
\left(\frac{\partial S_{\rm r\theta}}{\partial r}\right)^2
+\frac{1}{r^2} \left(\frac{\partial S_{\rm r\theta}}{\partial
\theta}\right)^2 +\nonumber \\ && \frac{1}{r^2\sin^2\theta}
\left[{\cal L}+\frac{eB}{2}r^2 \left(1+\frac{3M^2}{10{\omega}
r^4}\right)\sin^2\theta\right]^{2}.
\end{eqnarray}

\begin{figure*}
\includegraphics[width=0.45\textwidth]{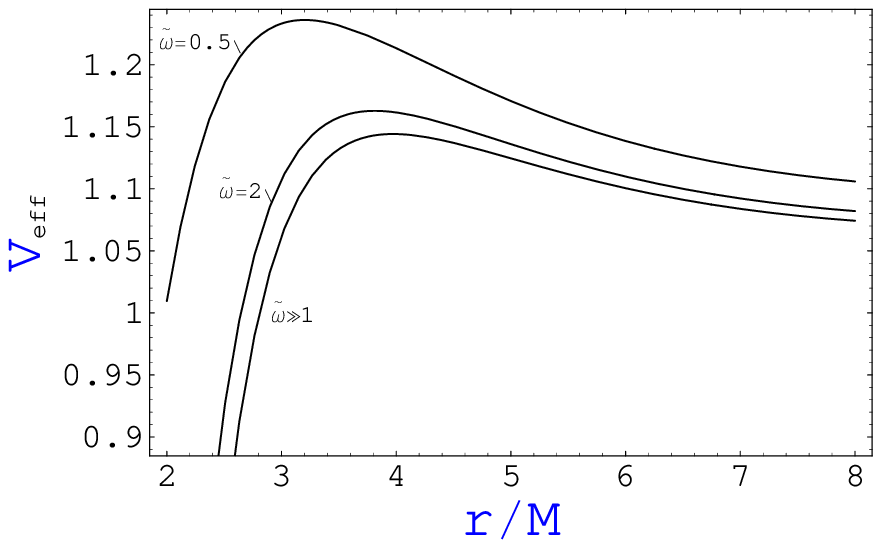}
\includegraphics[width=0.45\textwidth]{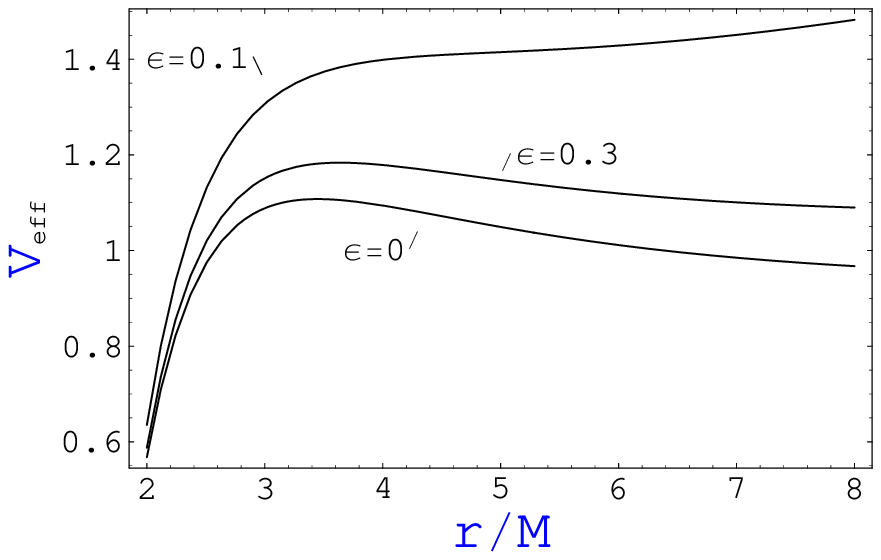}
\caption{\label{fig:1} {Radial dependence of the effective
potential of the radial motion of the charged particle around
black hole immersed in a uniform magnetic field in
Ho\v{r}ava-Lifshitz gravity. In the left graph the effective
potential of the radial motion of the charged particle around
black hole has been shown for the different values of the the
dimensionless parameter $\tilde{\omega}$, the values of the
momentum ${\cal L}/mM=4.3$ and the magnetic parameter
$\epsilon=0.03$ are fixed. For comparison we have also plotted
this dependence in the case of the Schwarzschild black hole,
corresponding to $\tilde{\omega}\gg1$. In the right graph the
effective potential of the radial motion of the charged particle
around black hole has been shown for the different values of the
the magnetic parameter $\epsilon$, the values of the momentum
${\cal L}/mM=4.3$ and the dimensionless parameter
$\tilde{\omega}=1$ are fixed. }}
\end{figure*}

One can easily separate variables in this equation in the
equatorial plane $\theta = \pi/2$ and obtain the equation for
radial motion
\begin{equation}\label{urdvij}
\left(\frac{dr}{ds}\right)^2={\cal E}^2-V_{\rm eff}({\cal
L},r,\epsilon,\omega)\ ,
\end{equation}
where $s$ is the proper time along the trajectory of a particle
and
\begin{eqnarray}\label{eff_potential}
&&V_{\rm eff}({\cal L},r,\epsilon,\omega)= \left(1+\omega
r^2-\sqrt{r(\omega^2 r^3+4\omega
M)}\right) \nonumber\\
&&\qquad \times \left\{ \left[ \frac{\cal
L}{r}+\left(\frac{r}{2M}+\frac{3M^3}{10\tilde{\omega}r^3}\right)\epsilon
\right]^2+1\right\}
\end{eqnarray}
can be interpreted as an effective potential of the radial motion,
where the dimensionless parameter $\tilde{\omega}=\omega M^2$ is
introduced. Here we have changed  ${\cal E}\rightarrow {\cal E}/m$
and ${\cal L}\rightarrow {\cal L}/m$. The effective potential
besides the energy, the angular momentum, the KS parameter
$\omega$ and the radius of the motion also depends on the
dimensionless parameter $ \epsilon={eBM}/{m} $ which characterizes
the relative influence of a uniform magnetic field on the motion
of the charged particles. In Fig.~\ref{fig:1} the radial
dependence of the effective potential of the radial motion of the
charged particle around black hole for the different values of the
dimensionless parameter $\tilde{\omega}=\omega M^2$ and magnetic
parameter $\epsilon$ are shown. One can easily see, that orbits of
the particles become more unstable with increasing of the
parameter $\epsilon$ (Similar results have been obtained in our
previous paper \cite{aak08}.). Influence of the dimensionless
parameter $\tilde{\omega}$ is sufficient near the black hole:
as it is seen from the figure the potential carries the repulsive
character. It means that the particle coming from infinity and
passing by the source will not be captured: it will be reflected
and will go to infinity again. The orbits start to be only
parabolic or hyperbolic and no more circular or elliptical orbits
exist with decreasing the dimensionless parameter
$\tilde{\omega}$, i.e. captured particles by central object are
going to leave the black hole.
%

\section{\label{sec:4} Circular orbits around black hole in Ho\v{r}ava-Lifshitz gravity}

In order to find  solution for the ISCO radius of $r_{\rm ISCO}$
we assume that the external magnetic field is absent.

\begin{figure*}
\includegraphics[width=0.45\textwidth]{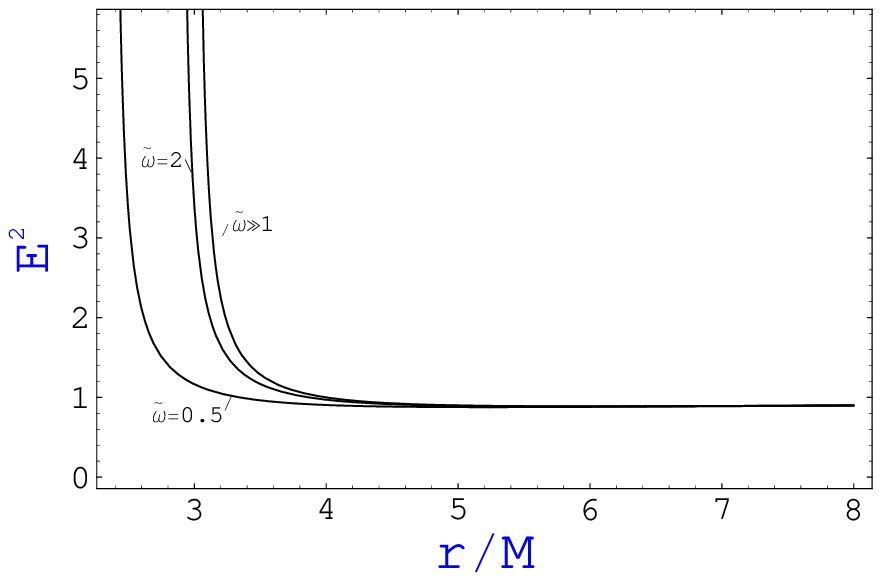}
\includegraphics[width=0.45\textwidth]{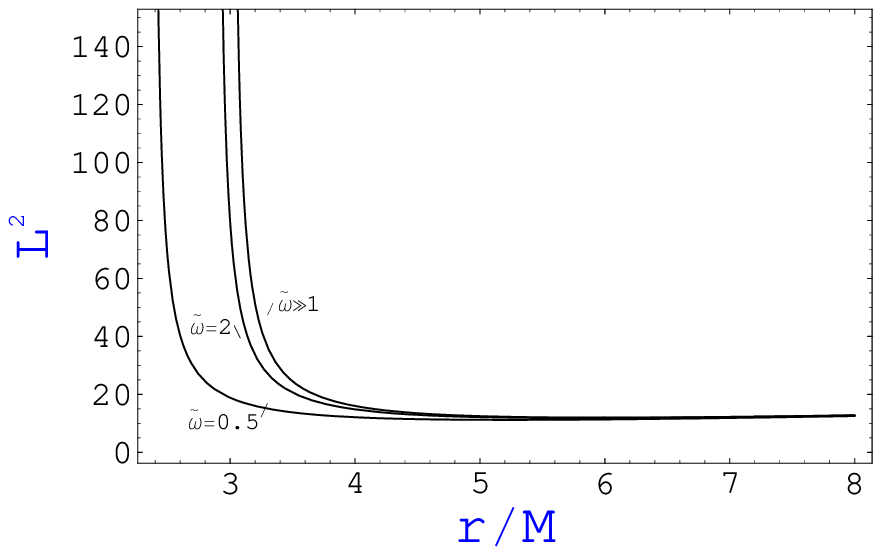}
\caption{\label{fig:2} {Radial dependence of energy (left graph)
and angular momentum (right graph) of the circular orbits around
black hole in Ho\v{r}ava-Lifshitz gravity for the different values
of the dimensionless parameter  $\tilde{\omega}$. For comparison
we have also plotted this dependence in the case of the
Schwarzschild black hole, corresponding to $\tilde{\omega}\gg1$.}}
\end{figure*}

The expression (\ref{urdvij})  can now be written as
\begin{eqnarray} \label{ruch}&& \left(\frac{dr}{ds}\right)^2
=f(r)=\\&&{\cal E}^2-\left(1+\omega r^2-\sqrt{r(\omega^2
r^3+4\omega M)}\right)\left( \frac{{\cal
L}^2}{r^2}+1\right).\nonumber
\end{eqnarray}
Using the equation (\ref{ruch}) and the condition of occurrence of
circular orbits:
$
f(r)=0\,,\ f'(r)=0\ ,
$
one can easily find  expressions for energy ${\cal E}$ and angular
momentum ${\cal L}$ of a circular orbit of radius $r_{c}$, which
are given as
\begin{eqnarray}
&& \hspace{-0.4cm}\label{encirc}{\cal E}^2=
\left(1+\frac{M-r^3(\Sigma-1)\omega}{3M
-r\Sigma}\right) \left(1-r^2\omega(\Sigma-1)\right)  ,\\
&& \hspace{-0.4cm} \label{enmomn}{\cal L}^2= r^2\frac{M-\omega
r^3(\Sigma-1)}{3M-r\Sigma} \ ,
\end{eqnarray}
where notation $\Sigma=(1+4M/r^3\omega)^{1/2}$ has been used.
Fig. \ref{fig:2} shows the radial dependence of both the energy
and the angular momenta of the test particle moving on circular
orbits in the equatorial plane. One can easily see that circular
orbits corresponding to constant value of the energy and momentum
of the test particle shift to the central object with decreasing
of the parameter $\tilde{\omega}$.
\begin{table*}
\caption{\label{1tab} The innermost stable circular orbits around
black hole and the critic values of the momentum of the particles
falling down to the central black hole in Ho\v{r}ava-Lifshitz
gravity.}
\begin{ruledtabular}
\begin{tabular}{cccccccccc}
 $\tilde{\omega}$ & 0.5 & 1 &  2 &  4 &  6 & 8 & 10 & 12\\
\hline
$r_{\rm ISCO}$ & 5.23655 & 5.66395 & 5.84024 & 5.92193 & 5.94834 &
5.9614 & 5.96918 & 5.97436
\\
\hline
${\cal L}_{\rm cr}^2$ & 14.77 & 15.454 & 15.7395 & 15.8725 &
15.9156 & 15.9369 & 15.9496 & 15.9581
\\
\end{tabular}
\end{ruledtabular}
\end{table*}

{For the existing circular orbits the expression for angular
momentum (\ref{enmomn}) of the test particle requires, in
particular, that $r\Sigma-3M \geq 0$.
Consequently,} one can easily find minimum radius for circular
orbits $\tilde{r}_{\rm mc}={r}_{\rm mc}/M$:
\begin{equation}\label{orgeq}
\tilde{r}_{\rm mc}=\left\{\begin{array}{rccl} \frac{3
\tilde{\omega}}{(\Delta-2\tilde{\omega}^2)^{1/3}}+\frac{(\Delta-2
\tilde{\omega}^2)^{1/3}}{\tilde{\omega}}\ ,& \tilde{\omega}<\frac{2\sqrt{3}}{9}\\
\\
2\sqrt{3}\cos\left[\frac{1}{3}\arccos\left(-\frac{2 \sqrt{3}}{9}
\frac{1}{\tilde{\omega}}\right)\right]\ , & \tilde{\omega}\geq
\frac{2\sqrt{3}}{9} \end{array}\right.
\end{equation}
where $\Delta=(4 \tilde{\omega}^4-27 \tilde{\omega}^6)^{1/2}$. The
obtained equation (\ref{orgeq}) is the original one.

As it was expected in the limiting case when $\tilde{\omega}\gg
1$, i.e. when metric (\ref{metric}) coincides with the well known
Schwarzschild metric, one can easily obtain the known result for
minimal radius of circular orbits around the Schwarzschild black
hole as $\tilde{r}_{\rm mc}=3$.

\begin{figure}
\includegraphics[width=0.45\textwidth]{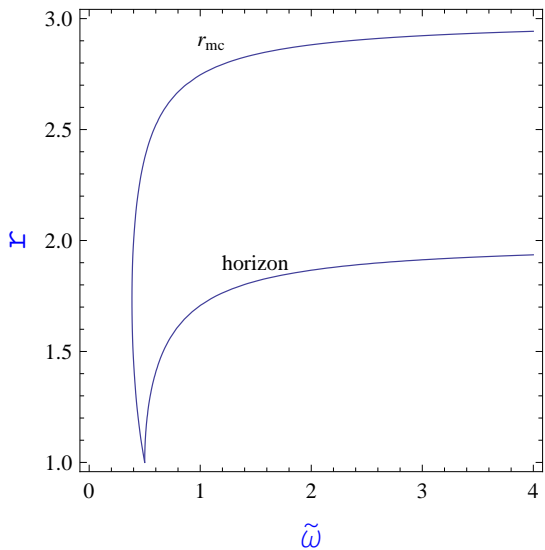}

\caption{\label{fig:romg} {Dependence of the radius of the horizon
and $r_{\rm mc}$ from the $\tilde{\omega}$.  }}
\end{figure}

{In the fig. \ref{fig:romg} the dependence of the both radii of
the horizon and $r_{mc}$ from dimensionless parameter
$\tilde{\omega}$ are shown. In the case of $\omega \geq
{2\sqrt{3}}/{9} $ the decreasing of the KS parameter
$\tilde{\omega}$  forces the minimum radii of the circuar orbits
$r_{\rm mc}$ to shift to the central object. In the case of
$\omega < {2\sqrt{3}}/{9}$ there is no lower limit for $r_{\rm
mc}$, which means that circular orbits can be present near the
black hole.}

The minimum radius for a stable circular orbit will occur at point
of inflexion of the function $f(r)$, or in other words where the
supplement conditions $f(r)=f'(r)=0$ with the relation $f''(r)\geq
0$ are satisfied. The numerical results for the values of ISCO
radii for the different values of the parameter $\tilde{\omega}$ {
in the case of $\tilde{\omega}\geq 0.5$} are given in the table
\ref{1tab} (the second line). {From the results one can easily get
in the limit of Schwarzschild spacetime $\omega r^{2}\rightarrow
\infty $ the standard value for ISCO radius as $r_{\rm ISCO}=6M$.
}

Next, we will consider the capture cross section of slowly moving
test particles by black hole in Ho\v{r}ava-Lifshitz gravity (Slow
motion means that ${\cal E}\simeq 1$ at the infinity.).  The
critical value of the particle's angular momentum, ${\cal L}_{\rm
cr}$, hinges upon the existence of a multipole root of the
function $f(r)$ in (\ref{ruch}) \cite{zakhar}.
We give the numerical results for ${\cal L}_{\rm cr}^2$ in the
Table \ref{1tab} (the third line).

{In the limiting case, i.e. when $\tilde{\omega}\gg1$ the value of
the critical angular momentum is ${\cal L}=4$, which coincides
with critical angular momentum for particle capture cross section
for the Schwarzschild black hole \cite{mtw}. As a particle having
the critical angular momentum travels from infinity towards the
black hole in Ho\v{r}ava-Lifshitz gravity, it spirals into an
unstable circular orbit.}

\section{\label{conclusion} Conclusion}

\begin{figure}
\includegraphics[width=0.45\textwidth]{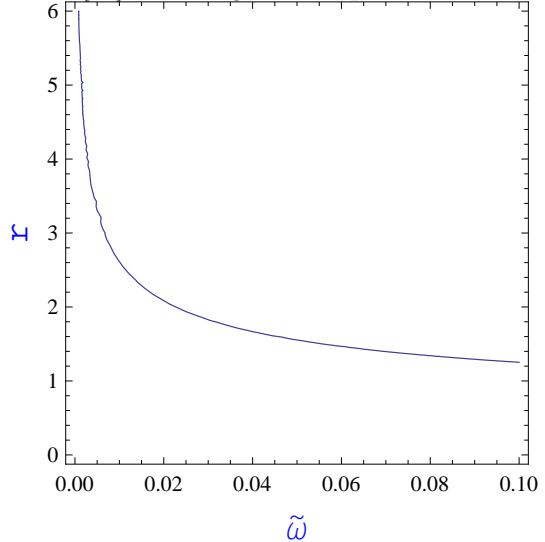}

\caption{\label{iscoomg} {Dependence of the ISCO radius from the
$\tilde{\omega}$.}}
\end{figure}

{Constraints for the KS parameter from the Solar system tests were
found as $(5.660 \pm 3.1) \cdot 10^{-26} {\rm
cm}^{-2}$~\cite{lobo1}. The similar constraints for the parameter
$\omega\simeq 1.25 \cdot 10^{-25} {\rm cm}^{-2}$ have been found
from the quantum interference effects~\cite{htaa}.}
{In the fig. \ref{iscoomg} the dependence of the $r_{\rm ISCO}$
from dimensionless KS parameter $\tilde{\omega}$ is shown. From
figure one can see that in presence of the parameter
$\tilde{\omega}$ ISCO shifts to the central black hole.}
One can easily compare the obtained numerical results with
observational data for ISCO radius for some candidates of rotating
black holes~\cite{narayan}.
One can obtain the lower value for the parameter as $\omega \simeq
3.6\cdot 10^{-24} {\rm cm}^{-2}$.

\acknowledgments

This research is supported in part by the UzFFR (projects 1-10 and
11-10), projects FA-F2-F079 and FA-F2-F061 of the UzAS. The
authors acknowledge the hospitality at the IUCAA, Pune, India.
Authors would like to thank the anonymous Referees for very useful
comments and careful reading of the paper.

\end{document}